# Special Relativity and Time Travel Revisited

Akhila Raman
Berkeley, CA.
Send comments to:  akhila_raman@yahoo.com
Jan 23, 2001.

**Abstract:**

In this paper, Lorentz Transformation(LT) is derived by an alternate method, using photon clocks, placed at the locations of the concerned events, which are initially synchronised using a light signal(Einstein synchrony). Then, it is shown that the second term in the time dilation equation of the LT, is the term responsible for time travel and further it is shown that this term arises merely due to **non-simultaneous initial clock synchronisation** and that this term does not correspond to any actual time elapsed while timing events, thus **ruling out the possibility of time travel**, from the framework of special relativity. It is also shown that only the first term in the time dilation equation of LT represents the true time dilation. In other words, it is argued that the time dilation equation of the LT, corresponds to the actual reading of the clock which is not necessarily the actual time elapsed while timing an event, the difference being attributed to the clock synchronisation mechanism.

## 1. Introduction

Special Relativity(SR) states that the physical laws of nature take the same form in all uniformly moving inertial frames and remain unchanged under a Lorentz transformation. It also states that light speed "c" is constant in all inertial frames and is independent of the speed of its source.

Let an event take place at a distance "X" and time "T" in one frame of reference F, the same event will have coordinates X' and T' in another frame F' moving at a relative velocity "v" w.r.to F, according to the following Lorentz Transformation(LT):[1]

$X' = ( X - v*T )/K$         --- eqn A
$T' = (T - X*v/c^2) /K$      --- eqn B
where   $K = \sqrt{1 - v^2/c^2}$

We can see that when X=c*T, X'= c*T'.
An implication of SR is that identical clocks in 2 inertial frames moving at a relative velocity will tick differently(time dilation). And distances will be measured differently.

These equations can be geometrically represented as in Fig 5. in Appendix. From eqn. B, we see that when **$X*v/c^2 > T$,** T' will have a sign opposite to that of T! This means that an event which is in the past of one frame, can appear in the future of another frame moving at a relative velocity w.r.to the former! Say, an observer O at rest and another observer O' in motion pass each other at the origin(T=0,T'=0). An event $E_r$ which is in the past of O(T<0) will be in the future of O'(T'>0) for the **region R with both T<0 and**

**T'>0 bounded by thick lines**, as depicted in Fig 5, thus allowing the possibility of time travel. However, this region R is always outside the light cone of both observers, meaning that neither do they know of the event E, nor can they influence the event, given nothing can travel faster than light(FTL)! BUT, using concepts like wormholes, theoretically one can travel "effectively" FTL and this introduces the disturbing possibility of time travel, causality violations and unsolvable paradoxes. For example, if one could go back into the past and kill one's own grandfather when he was a child, then how does one explain one's own existence?[2]

In short, the region R enables time travel and the term responsible for it is the second term in eqn B, namely E=**X*v/(K*c$^2$** ). If this term were not present, T' and T will have the same sign always, thus ruling out the possibility of time travel. (note that another region with both T>0 and T'<0 also allows time travel and similar arguments apply).

## 2. Derivation of Lorentz Transformation:

Usually, LT is derived by considering the trajectory of a photon in the 2 inertial frames and applying relativity postulates and boundary conditions[5]. This method does not give an intuitive feel of why a clock has to tick at different rates inherently between moving frames. Hence, I have made an attempt to derive it in a different manner using fixed and moving photon clocks placed at the location of the concerned events. This method requires that all clocks in the 2 frames be synchronised by,say, sending light signals from a midpoint between any 2 clocks in a given frame, before using them to time events.

This method consists of 3 parts:
1. Time dilation: How moving clocks tick differently
2. Initial Clock Synchronisation and Relativity of Simultaneity
3. Computing the time difference between 2 events separated by a distance.

Let us consider 2 events E1 and E2 which take place at points A and B respectively,which are separated by a distance X and time T ,in frame F. We wish to compute corresponding distance X' and time T', as measured in F'.

### 2.1 Time dilation: How moving clocks tick differently

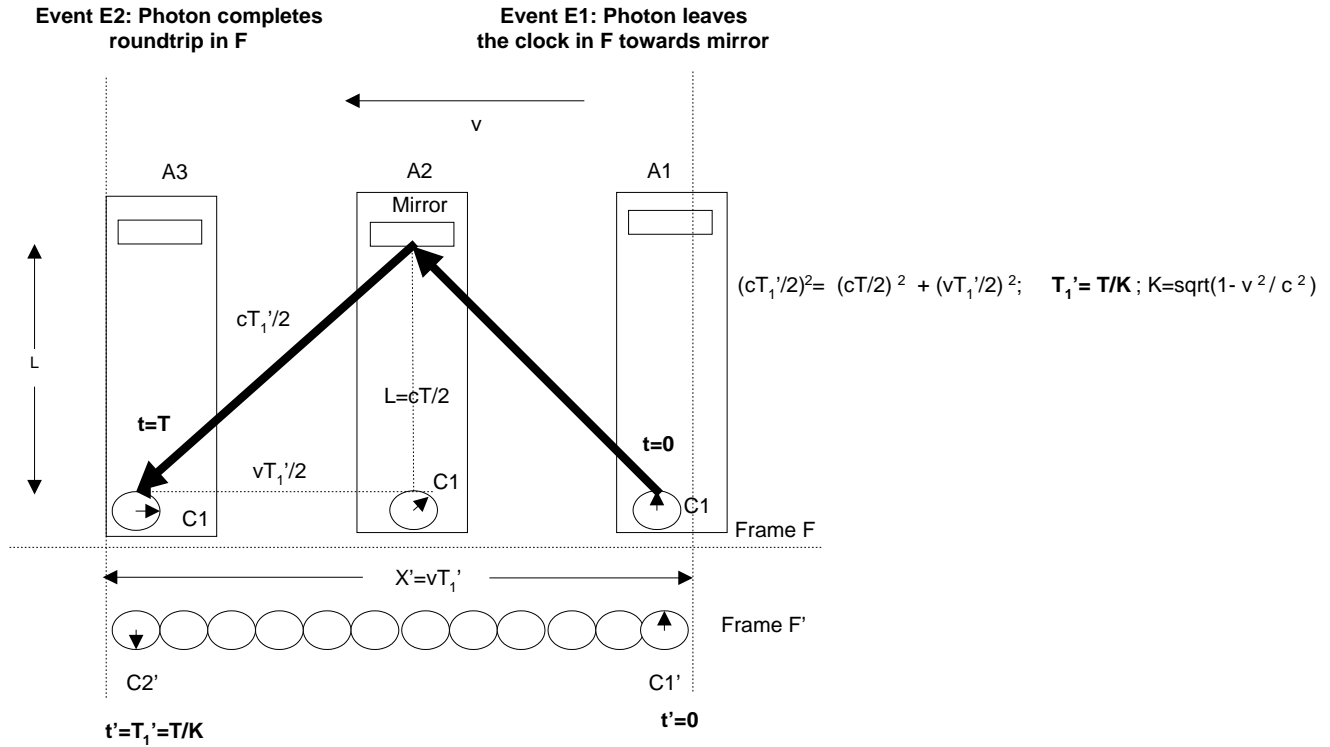

**Fig 1.** Diagram depicting **time dilation**; We wish to compare the relative times of clocks between 2 inertial frames with relative velocity "v". From the viewpoint of Frame F', F' is at rest and F is moving with velocity "v". The clock apparatus consists of a photon travelling from a source located beside a regular clock C1 to a mirror and travels back, as shown in the figure for frame F. A similar apparatus exists for frame F', but for sake of simplicity, only clocks are shown(e.g. C1', C2' ). An array of identical clock apparatus is placed along the length of horizontal axis in F'. From viewpoint of F', a similar array of apparatus in F are moving to the left, of which only one moving apparatus is shown. A1,A2,A3 represent successive states of the same moving apparatus as it moves left w.r.to frame F' with relative velocity "v". When the photon completes a roundtrip in frame F, we wish to compare the relative times. We can see that as the photon travels towards the mirror in F, the apparatus A1 has moved left w.r.to F'. But the photon speed is constant in all inertial frames, hence the photon in F would take longer to complete a roundtrip and by that time, the clock apparatus in F' would show a time increased by a factor K, by a simple application of pythogorean theorem. In summary, any clock which is present at both events measures proper time and shows the minimum time,as in the moving clock C1 in this example; **Two clocks separated by a distance measure improper times between 2 events and will show a time increased by a factor K** as in the case of the clocks C1' and C2' in frame F' in this example.

This is well explained in Fig.1 above.[1] A photon clock apparatus is a simple setup in which a photon travels from a source to a mirror and then back and everytime it completes a roundtrip, it illuminates the clock C1 beside the source. A similar array of apparatus is present in frame F' also along the horizontal axis, only the clock values are shown for simplicity. We wish to compare the relative clock times in the 2 frames, upon completion of a roundtrip in frame F. From the viewpoint of an observer O' in F', frame F is moving to the right with velocity "v**", he sees the photon travelling a longer diagonal distance to complete a roundtrip** and given the fact that photon speed is the same in all inertial frames, the roundtrip will take longer in F and by that time clock in F' will show a time greater by a factor of K(using simple pythogorean theorem).

The keypoint is that: **any clock which is present in both events**(clock C1 in this case, the events being sending and receipt of the photon) shows the minimum possible time which is termed as the "**proper time**". **Two different clocks** spaced apart measuring spatially separated events, will show a time **greater by a factor of K** and is termed as "**improper time**".(**Statement A**)

Note that this time dilation applies not only to photon clocks, but also to any mechanical clock and our biological clocks!

## 2.2 Initial Clock Synchronisation and Relativity of Simultaneity:

It is impossible to synchronise the clocks at points A and B in both the frames(totally 4 clocks for a pair of points) to zero at the same time. This is because one of the frames is moving w.r.to the other and hence light signal in the moving frame reaches one of the clocks earlier and the other clock later, while the signal reaches both clocks at the same time in the rest frame. Hence, out of the 4 clocks, while 3 read zero, the fourth will lead or lag by a value E= $X*v/(c^2 *K)$, depending on the direction of motion. This is clearly depicted in Fig 2.a through Fig 2.d in Appendix. Hence clocks which are simultaneous in one frame, are NOT simultaneous in another frame, and this is termed as **Relativity of Simultaneity**, as depicted in Fig 2.e.

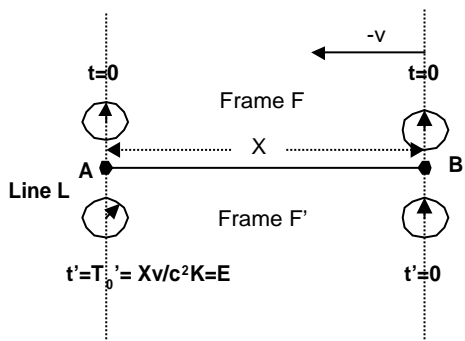

**Fig 2.e.** Diagrams showing the relativity of simultaneity

Note that whichever method of synchronisation one chooses, whether using light signals, or by slow clock transport(synchronising them all at one location and then slowly transporting them to the desired location by say walk), this term E will be always present.[4]

It is interesting to note that there is no previleged frame of reference. Each frame might think it is at rest and the other frame is moving and there is no absolute validity to both claims in inertial frames. Hence when an observer O in frame F, located at point A, sees that clocks read zero at point A in both frames($T_1$, $T_1$'), cannot say with absolute confidence that the clock at point B in his frame($T_2$) reads zero "at the same time" as when his clock at A($T1$) reads zero, simply because there is no way he can be sure whether his inertial frame is at rest or moving! However he can confidently say that **when** his clock at B($T_2$) reads zero, other frame's clock at B ($T_2$')  leads/lags by   "E" ! And the same goes for the observer in frame F'.

Thus we see that, there is a **residual synchronisation offset factor of E** , between corresponding pairs of clocks in F and F'.[3]

## 2.3 Computing the time difference "T" between 2 events separated by a distance "X":

If we simply take the difference of the 2 clock values at points A and B, after synchronisation, we are assuming that the initial start value of both the clocks is zero. This assumption is OK for measurements within a frame, BUT when we ask the question: When the time difference between 2 events is "T" in frame F, **what is the corresponding time difference in F'**, it would be wrong to merely take the difference of clock readings in F' also, because we already know that **the clocks in F' did not both read zero at the instant when clocks in F started with zero values and that one of them was offset by a factor E**! The correct method would be to **subtract the corresponding start values of clocks in F' from the clock readings in both locations** and **then** take the difference. This is my main argument.

Let us derive the LT by using clocks as mentioned above. The method is detailed in Fig.4.a in Appendix. First we synchronise the 4 clocks(for a pair of points A and B) as detailed in Sec 2.1 and as depicted below in Fig 4.b. So when the clocks in F read zero at points A and B, corresponding clocks in F' read "E" and "0". Now let an event E1 take place at A(x=0,t=0) and another event E2 take place at B(x=X,t=T). Let us insert a dummy event E0 at B(x=X,t=0).

Event E1: $t=0$, $t'=E$ (due to relativity of simultaneity)
Event E0: $t=0$, $t'=0$
Event E2: $t=T$, $t'=T/K$ (time dilation as detailed in Sec 2.1)

Time difference between E1 and E2 in frame F= $T_d$ = $T - 0 = T$
Time difference between E1 and E2 in frame F'= $T_d'$ = $T/K - E = (T - X*v/c^2)/K$
                         ---eqn C

Thus we have derived the time dilation equation of LT as in eqn B! Note that we have merely read off the clock values and taken their difference, which assumes ALL the clocks started timing events from zero start value, which is not correct!

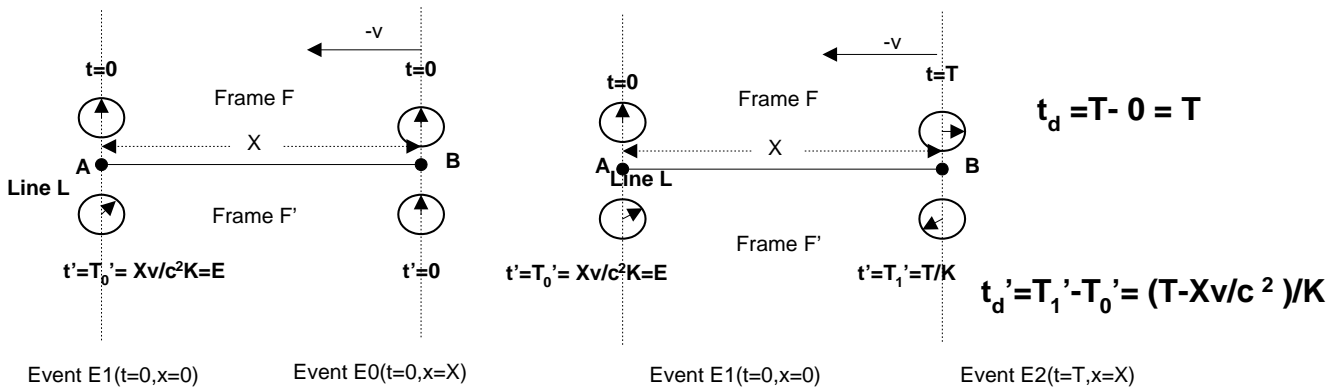

**Fig 4.b**. Diagrams showing the readings of the clocks in frame F and F' during events E1,E0 and E2,depicting time dilation; Length dilation is not shown

**Correct method of computing time difference:**

Time difference between E1 and E2 in frame F= $T_d$ =  (T –0)- (0-0) = T
**Corresponding time difference** between E1 and E2 in F'=
$$T_{cd}' = (T/K-0)- (E - E) = T/K \quad \text{--- eqn D}$$

Thus we see that, when the time difference between 2 events in frame F is "T", corresponding time difference in F' is only T/K.
Now, we see that both T and $T_{cd}'$ have the same sign and hence time travel is ruled out!

LT is still valid, except that **T' in LT denotes the actual reading of clocks** and their difference, rather than **the corresponding** time elapsed between 2 events. This is the key difference in **interpreting T'** in Lorentz Transformation. Time dilation term T/K still exists, but the time travel term E arises due to non-simultaneous clock synchronisation, not actual time elapsed while timing an event!

Similarly eqn A can be derived for length dilation as depicted in Fig 4.a. and Fig 3 in Appendix.

**Points to remember:**

a. Constancy of speed of light in all inertial frames is still preserved. Because any experiment done to compute light speed in either of the frames is an internal experiment and uses the reading of the clock as a time measure which is perfectly OK for that frame and since we have not altered LT equations, we still have X'=c*T' when X=c*T.

b. Especially those events for which **$X*v/c^2 > T$,** they are not causally connected, (one event could not have caused the other causally because X/c >T) , and it is not possible to determine which event occurred first and which one occurred later and reckon which event is in the past or future of the other. In such cases, we simply subtract the clock readings off, say, the synchronisation start values namely zero, at both locations in one frame and then take the difference and depending on the sign of the difference , we reckon about which event precedes the other. This reckoning is arbitrary, due to the residual uncertainty factor "E" during clock synchronisation(since neither can be absolutely sure who is at rest and who is in motion). But it does not matter because those events are not causally connected anyway! The only thing that matters is: we take the **corresponding time difference** in the other frame, and then we will see that reckoning of past and future in both frames are consistent, since they differ only by a factor K.

In addition, when an event happens, relatively moving observers belonging to different frames, passing by that event and coinciding with it at the moment it happens, will disagree about the location and time of that event, simply because their clocks ticked at different rates and their distance measures were different. But they will agree on the fact that the event actually took place and that the event is an absolute physical reality.

c. In Section 2.1, there is no reason why frame F' is previleged to be at rest. It is quite possible that F' is actually moving and F is at rest! In this case, clocks in F' will tick slower, and **upon completion of roundtrip of photon in F, the clocks in F and F' will still show the same values**! BUT, **the actual time elapsed** in F' between the clocks $C_1$' and $C_2$' = $T_1$' = $T/K - E_1$ where

   clock synchronisation residual offset factor between $C_1$' and $C_2$' =
   $E_1 = X*v/(c^2*K) = (vT) * v/(c^2*K) = T/K * (v^2/c^2)$.
   [Note: $X = vT$ from viewpoint of frame F in Sec 2.1]

   Hence $T_1$' = $(T/K) * (1 - (v^2/c^2)) = (T/K) * K^2 = T*K$.

In either case, time dilation is a true phenomenon, not merely an observational effect. Depending on who is moving and who is at rest, the actual time elapsed is greater or smaller by a factor K. However, the statement A given in section 2.1 is still valid, because it pertains only to times **shown** by clocks, not to **actual time elapsed** ! Infact, actual time elapsed cannot be determined in absolute sense in inertial motion. But, we can confidently say that the actual time elapsed between 2 clocks measuring improper time **differs** from the proper time by factor K. We do not know if it is greater or smaller. In the case of twin paradox, the physical asymmetrical aging between twins is a physically true phenomenon. Because one of the twins who went on a spacecraft experiences acceleration(non-inertial motion) on turning back to earth and hence both know who was at rest and who was moving and there is no ambiguity about actual time elapsed. Hence both can agree that the twin on spacecraft aged slower than the twin on earth at rest.

**Conclusion:**

We have derived Lorentz Transformation by an alternate method by using photon clocks positioned at the location of concerned events. It is argued that the parameter T' in LT represents the actual reading of the clock and **not** necessarily the actual time elapsed while timing an event, the difference attributed to the initial clock synchronisation offset due to **non-simultaneous synchronisation**. It is proposed that the actual time elapsed between 2 spatially separated events should be computed with reference to the corresponding clock synchronisation start values at respective locations and then taking the difference and it is shown that T' computed this way differs from T only by a factor K. Since both T' and T have the same sign, **possibility of time travel is ruled out**, from the framework of special relativity.

**References:**

1. James H. Smith "Introduction to Special Relativity" 1965.

# Appendix: Diagrams depicting derivation of Lorentz Transformation

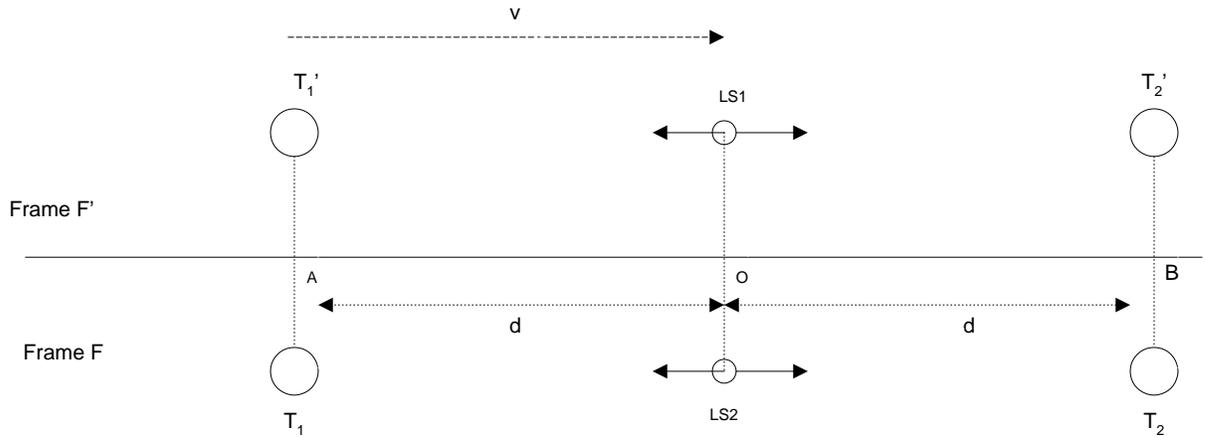

**Fig. 2.a**. At the same time t=0, light sources LS1 and LS2 at O(x=0) emit a synchronizing light signal towards A and B

At that instant, clocks $T_1'$ and $T_1$ coincide at point A(x=-d); clocks $T_2'$ and $T_2$ coincide at point B(x=+d);Distance AB=X=2d

Clock $T_1$ and $T_2$ in Frame F are at rest; In frame F', clock $T_1'$ is headed towards LS1 and clock $T_2'$ is headed away from LS1

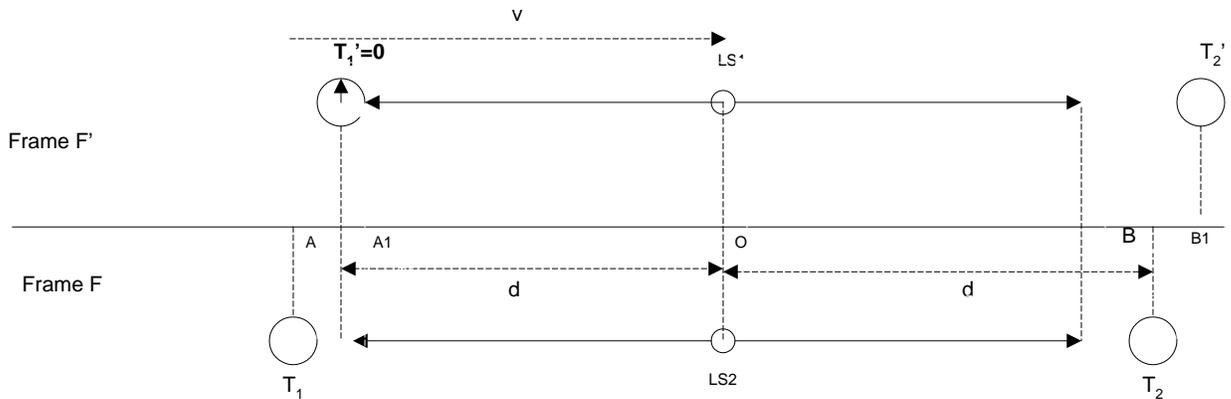

**Fig. 2.b**. At time $t=t_1=d/(c+v)$, light signal reaches $T_1'$ at point A1 and clock $T_1'$ is reset and starts ticking; distance $OA1=d- v*t_1$; $OB1=d+v*t_1$;

At that instant, clock $T_2'$ has moved away to point B1

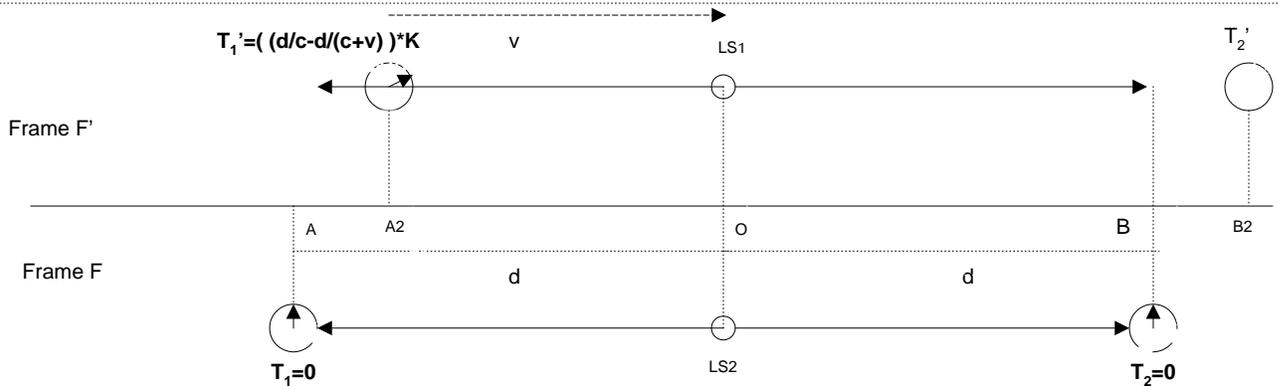

**Fig. 2.c**. At time $t=t_0=d/c$, light signal reaches $T_1$ at point A and $T_2$ at B and resets them and they start ticking; distance $OA2=d- v*t_0$; $OB2=d+v*t_0$; $T_1'=(t_0-t_1)*K$ since it measures proper time and hence multiplied by K.

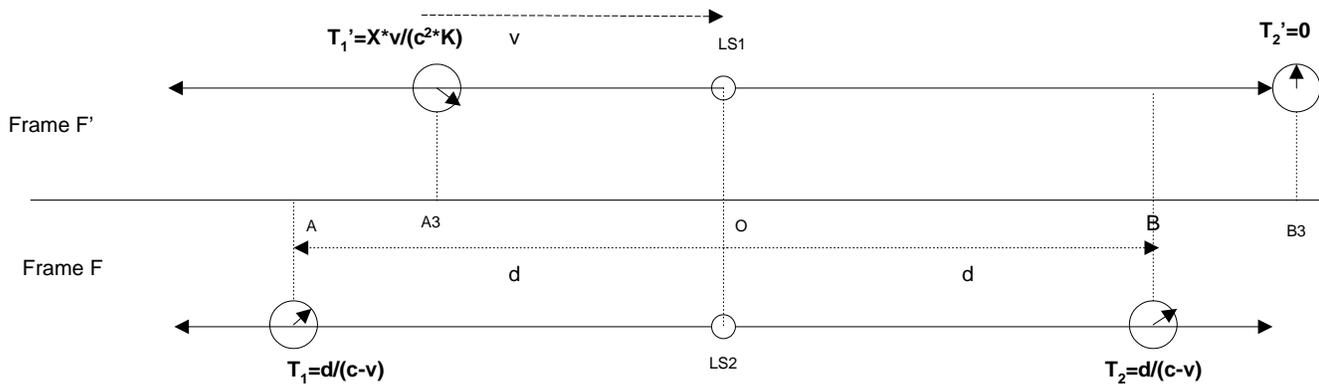

**Fig. 2.d.** At time $t=t_2=d/(c-v)$, light signal reaches $T_2'$ at point B3 and resets $T_2'$ to zero and it starts ticking; distance $OA3=d-v*t_2$; $OB3=d+v*t_2$; $T_1'=(t_2-t_1)*K=((d/(c-v)-d/(c+v))*K = X*v/(c^2*K) = E$; $T_2'=0$; But both $T_1=T_2=d/(c-v)$; **Thus we see that when 2 clocks in frame F are simultaneous, clocks in F' are offset in time by E, for a given distance "X" between any 2 clocks, due to the fact that synchronizing light signal reached them at different times.**

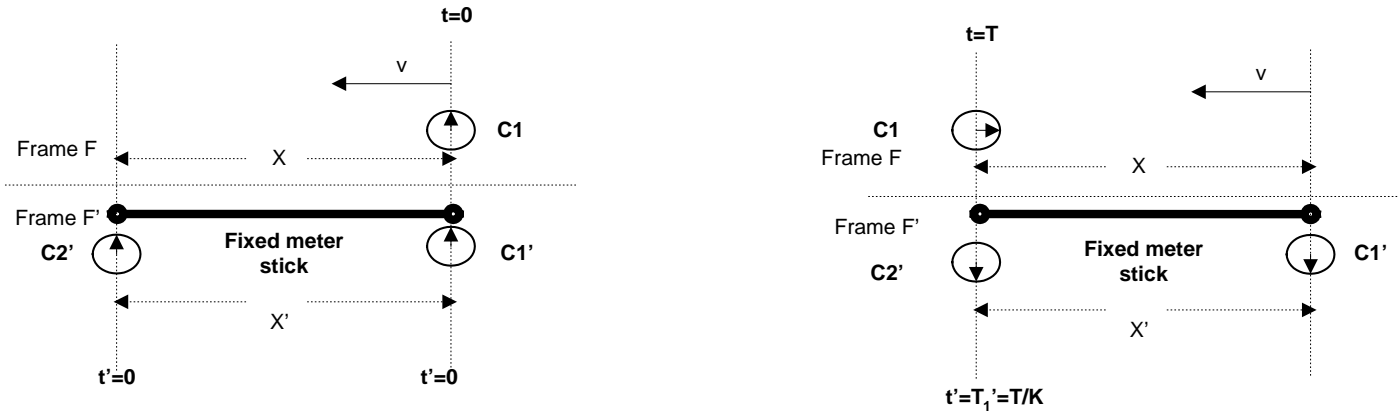

**Event E1:** From viewpoint of F', Moving Reference clock C1 in F passes one end of meter stick fixed in frame F', and is started on this event. At this instant, clocks in F' namely C1' and C2' are also started.

**Event E2:** From viewpoint of F', Moving clock C1 in F passes other end of meter stick fixed in frame F', and is stopped on this event. At this instant, clocks in F' namely C1' and C2' are also stopped.

**From viewpoint of F: Length of meter stick = $X = v*T$**
**From viewpoint of F': Length of meter stick = $X' = v*T_1' = v*T/K = X/K$ !**

**Fig 3.a.:** Diagram depicting Length dilation; The length of the moving meter stick is computed in both frames by measuring the **time elapsed between the passage of its two ends past the reference clock C1 in F,** and multiplying this by the known relative velocity "v". **From viewpoint of F**, his clock C1 is fixed and meter stick in F' is moving! He computes the stick length by multiplying the time elapsed in his clock C1 between the passage of the 2 ends of stick past C1, by "v".
**From viewpoint of F'** however, his meter stick is fixed and clock C1 is moving! He computes stick length by multiplying by "v", the time elapsed between his clocks C1' and C2', between the instants of passage of C1 past the 2 ends of the stick! Thus **the length "X" measured with a single clock C1 which measures proper time will be minimum**, while the **length X' measured with 2 clocks C1' and C2' which measure improper times will be greater by a factor of K**

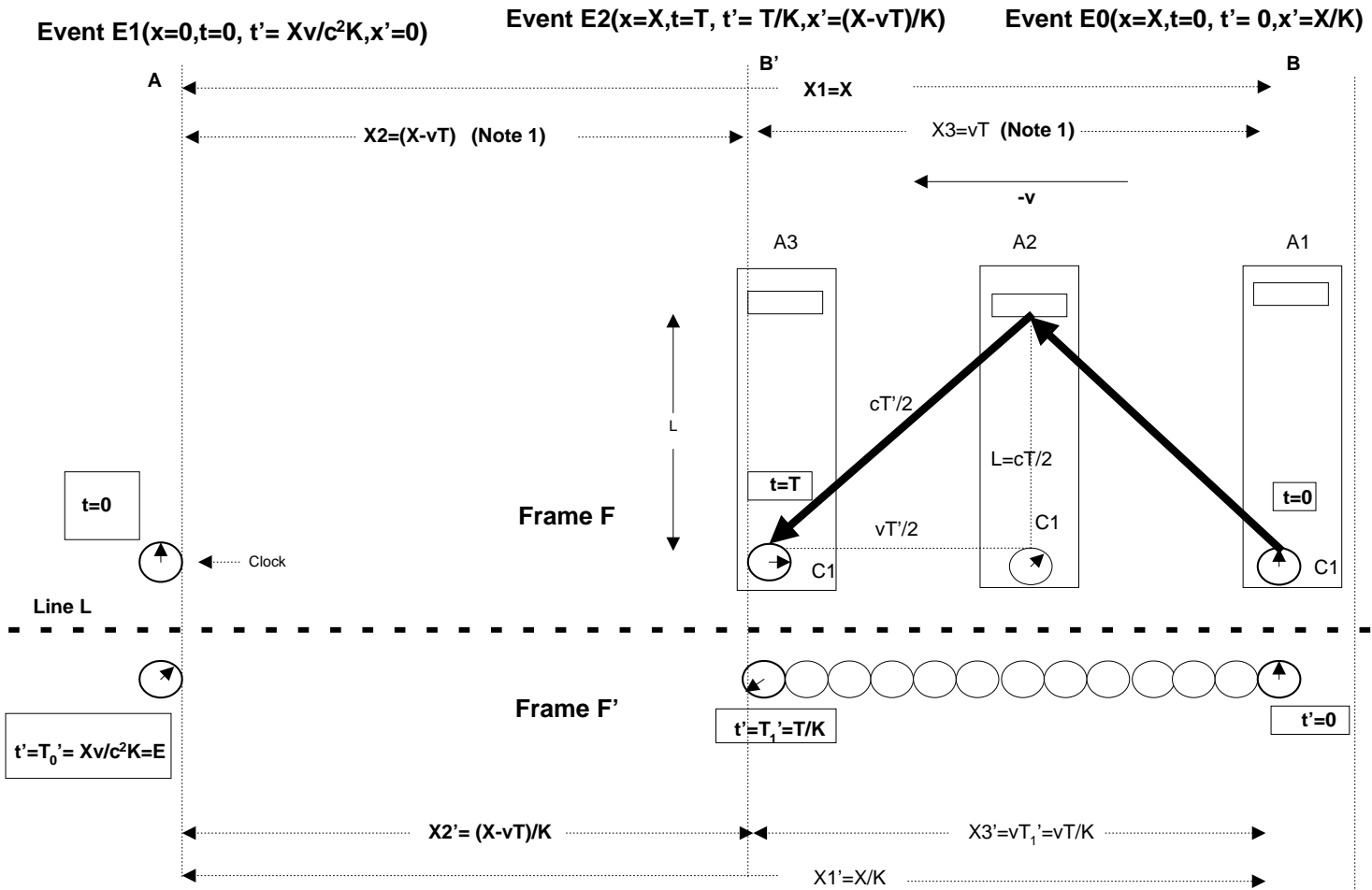

**Fig 4.a.** Inertial frame F is shown above line L with clocks positioned at specified distances, and frame F' is shown below L. From the viewpoint of an observer in frame F', F' is at rest and F is moving with relative velocity "-v"; A1,A2,A3 represent the successive states of the moving clock C1, from viewpoint of F'. However, from viewpoint of F, clock C1 is at rest at position B, B=B', and the entire clock array in F' is moving!

Time interval between events E1 and E2, in frame F = $t_d$ = T-0 = T

Time interval between events E1 and E2, in frame F' = $t_d'$ = $T_1'$ - $T_0'$ = $(T-Xv/c^2)/K$

Event **E1** happens at **t=0**, x=0 at frame F. Event **E2** happens at x=X, **t=T** in frame F. **A dummy event E0 simultaneous to E1, w.r.to frame F** is inserted at x=X, t=0; **Note that E0 and E1 are NOT simultaneous w.r. to F' and are offset by E, due to initial clock synchronisation, as explained in Fig 2;** When **E1** happens, reading of clock in F' = $T_0'$=E; Reading of clock in F' when **E2** happens = $T_1'$=T/K, due to time dilation as explained in Fig.1; E2 happens at the same location as E0 w.r.to frame F, but is seen to occur at a different location w.r.to frame F' due to relative motion. **Thus we see that 2 events E1 and E2 separated by a distance 'X' and time $t_d$ = T - 0 = T in frame F are separated by a time interval $t_d'$ = $T_1'$ - $T_0'$ = $(T-Xv/c^2)/K$ in frame F', thus proving the time dilation equation of Lorentz transformation**

**Note** that distances **X1,X2,X3** represent distances between events E1 and E0, E1 and E2, and E0 and E2 respectively, **projected on frame F', as judged by an observer in frame F,** who believes F is fixed and F' is moving and clocks these distances with clocks which measure proper times, as explained in Fig.3.

**(Ofcourse distance between E1 and E2= distance between E1 and E0 = X, projected on frame F, as judged by an observer in F.)**

**Note 1 ->** X1',X2',X3' represent distances between corresponding events **projected on frame F', as judged by an observer in frame F'**.

Note that X1',X2',X3' differ from X1,X2,X3 by a factor K, because the former distances are clocked with clocks which measures improper times, as explained in Fig.3.

**Thus, we see that 2 events E1 and E2 separated by a distance "X" and time "T" in frame F, are separated by a distance of X'=(X-vT)/K in frame F', thus proving length dilation equation of Lorentz Transformation.**

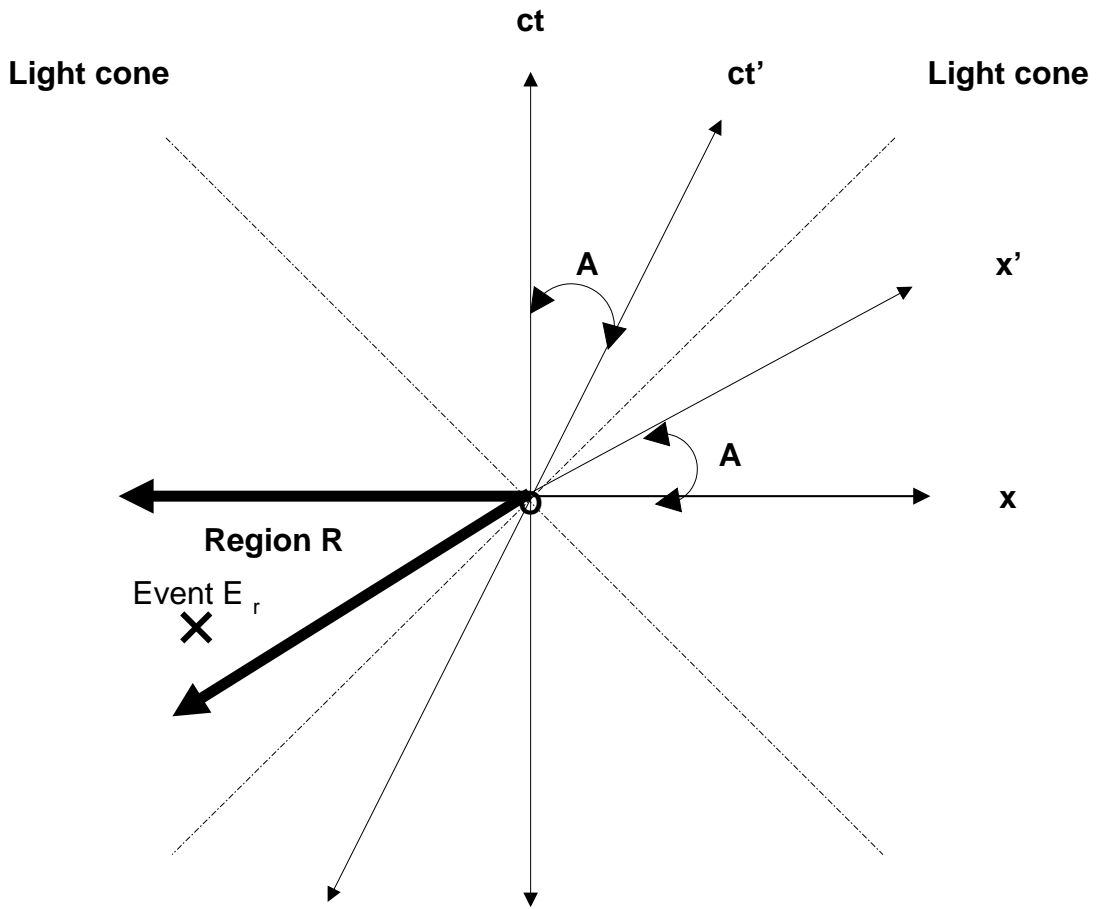

**Fig 5**: A= Angle between the 2 coordinate systems (x,t) and (x', t');
**tan A = v/c;**
Any event has coordinates (X,T) and (X',T') in the 2 systems which are related as:
**T' = (T - X*v/c$^2$ )/K**
**X' = (X - v*T) /K**
where K=sqrt(1 - v$^2$/c$^2$ )

These equations are called as **Lorentz Transformation** and the above diagram is their geometrical representation.

Note that the **event E$_r$ is outside the light cone**. When 2 observers with relative velocities pass each other at the origin, this event is outside their light cone and hence neither do they know of the event, nor can they influence this event.